\documentclass[12pt,preprint]{aastex}
\usepackage{amsmath}

\shorttitle{ULX}
\shortauthors{Ablimit at el.}

\begin{document}


\title{Formation and Evolution of Ultraluminous X-ray Pulsar Binaries to Pulsar-Neutron
Star and Pulsar-White Dwarf Systems}
\author{K. Abdusalam\altaffilmark{1}, Iminhaji Ablimit\altaffilmark{2,3,\star}, P. Hashim\altaffilmark{4}, G.-L L$\ddot{\rm u}$\altaffilmark{5,1},
  M. K. Mardini\altaffilmark{6,2}, and Z.-J Wang\altaffilmark{1}
}
\altaffiltext{1}{School of Physical Science and Technology, Xinjiang University, Urumqi 830046, China}
\altaffiltext{2}{Key Laboratory for Optical Astronomy, National Astronomical Observatories, Chinese Academy of Sciences, Beijing 100012,
China; iminhaji@nao.cas.cn}
\altaffiltext{3}{Department of Astronomy, Kyoto University, Kitashirakawa-Oiwake-cho, Sakyo-ku, Kyoto 606-8502, Japan; iminhaji@kusastro.kyoto-u.ac.jp}

\altaffiltext{4}{Astronomical Observatory of Xinjiang, Chinese Academy
of Sciences,  Urumqi 830046, China}

\altaffiltext{5}{Center for Theoretical Physics, Xinjiang University, Urumqi 830046, China; guolianglv@xao.ac.cn}

\altaffiltext{6}{Shandong Provincial Key Laboratory of Optical Astronomy and Solar-Terrestrial Environment,
Institute of Space Sciences, Shandong University, Weihai 264209, China}

\altaffiltext{}{$\star$ Corresponding Author}


\begin{abstract}

Recent observational and theoretical results have suggested that
some of ultraluminous X-ray (ULX) sources may contain neutron star (NS) accretors.
 However, the formation channel and properties of donor
stars of NS ULXs remain uncertain. By adopting the non-conservative
and rotation-dependent mass transfer model in the primordial binary
evolution, we investigate the way to form pulsar ULXs like observed
pulsar ULXs in a systematic way. Our simulation results indicate
that pulsar ULXs with Be stars and intermediate or/and high mass
donors match observed apparent luminosities, orbital periods and
observationally indicated donor masses of known pulsar ULXs. ULXs
with Be and intermediate donors are main contributors. The route of
accretion-induced collapse of WDs has 4.5\% contribution to the NS
ULXs, 4.0\% of NSs in ULXs are formed through electron-capture
supernovae (SNe), and 91.5\% of NSs in ULXs are born with
core-collapse SNe. We also studied the evolution of pulsar ULXs to
double compact star systems. We do not find NS-black hole systems
(merging in a Hubble time) that evolved from pulsar ULXs.
Pulsar-white dwarf (WD) cases that evolve through pulsar ULXs have
significant contributions to the whole NS-WD gravitational wave
sources. Contributions of pulsar-WD and pulsar-NS cases that
experienced pulsar ULXs are $\sim$40\% and 11\% among all LISA NS-WD
and NS-NS sources, respectively. Monte Carlo simulation noise with
different models give a non-negligible uncertainty.

\end{abstract}


\section{Introduction}

The Ultra-luminous X-ray sources (ULXs) were first discovered in
nearby galaxies (e.g., Fabbiano 1989), and ULXs are point-like
off-nuclear extragalactic sources with an X-ray luminosity of $
L_{\rm X} \geq 10^{39}\,\rm erg\,s^{-1}$ (see the review Kaaret et
al. 2017). The super- or sub-Eddington accretion onto stellar-mass
black holes (BHs) or intermediate mass BHs has been suggested as the
power sources of the high X-ray luminosity of ULXs (e.g. Colbert \&
Mushotzky 1999; Feng \& Soria 2011). Recently, pulsations in the
X-ray data of nine ULXs have been observed (see Table 1), and the
apparent X-ray luminosities in those ULXs are higher than the
Eddington limit for a typical neutron star (NS), and the
observed pulsations in these ULXs provided irrefutable evidence
that those ULXs contain accreting NSs (Bachetti et al. 2014;
F$\ddot{\rm u}$rst et al. 2016; Tsygankov et al. 2017; Israel et al.
2017; Carpano et al. 2018; Brightman et al. 2018; Sathyaprakash et
al. 2019).

The emission mechanism (producing high apparent luminosity) and
formation channel of ULXs are two main debates addressed in the last
decades. For the first argument, the presence of geometrically thick
accretion disks was invoked to produce the observed high
luminosities of ULXs by causing beaming emission (Shakura \& Sunyaev
1973; King et al. 2001; King 2009; Poutanen et al. 2007). The very
strong magnetic field ($B\sim10^{14}$ G) of the magnetar-like NS was
proposed as one possible physical mechanism to increase the
effective luminosity by reducing the electron scattering
cross-section (e.g., Herold 1979; Dall'Osso, Perna \& Stella 2015;
Mushtukov et al. 2015). However, King \& Lasota (2019) suggested
that the observed ULX properties are explained by NSs with normal
magnetic fields and not by the presence of magnetars, and Table 1 of
King \& Lasota (2020) shows derived magnetic field strengths of
pulsar ULXs are lower than that of a normal magnetar. Besides, the
highly super-Eddington accretion disc that extends close to the
accretor without the very strong magnetic field was studied to
generate the extreme luminosities (e.g. Klu$\acute{\rm z}$niak \&
Lasota 2015). For the observed correlation between pulse fraction
and X-ray photon energy in pulsar ULXs,  King \& Lasota (2020)
suggested that NS spin axes significantly misaligned from their
central accretion discs, and discussed that scattering in the
funnels collimating their emission and producing their apparent
super-Eddington luminosities is the most likely origin of the
correlation. To explore the exact emission process to produce
super-Eddington luminosities, more observational evidences and
investigations are required.

In the second argument, the nature of donor stars is the most
uncertain property of ULXs. Observationally, the far distances,
bright accretion disks, and the significant variations in the
spectra of ULXs from year to year make the donor stars and their
properties hard to detect (see Kaaret et al. (2017) and Heida et al.
(2019) for more discussions). There are rough mass estimations for
seven pulsar ULXs (see Table 1),
 and observational results indicate that
high mass or intermediate mass stars are more likely be donors in
ULXs. In the theoretical frameworks for the formation channel,
Fragos et al. (2015) investigated the origin of the NS
ULXs. They considered non-conservative mass transfer and specific
orbital angular momentum loss due to the mass loss in the population
synthesis simulation code BSE (Hurley et al. 2002) and detailed
binary calculation of the MESA code (Paxton et al. 2015). Initial
masses of $3-8\,M_\odot$ in relatively short orbital periods (1-3
days) have been suggested for NS ULXs by Fragos et al. (2015).
Subsequently, the binary population synthesis (BPS) study of
Wiktorowicz et al. (2017) demonstrated that NS X-ray binaries may
significantly contribute to the ULX population compared to BH X-ray
binaries (see also Shao \& Li (2015)), and showed that donor stars
in NS ULXs tend to have lower masses ($<2\,M_\odot$). BPS results of
Wiktorowicz et al. (2019) found that most BH ULXs have isotropic
X-ray emissions, while beaming X-rays are more common in NS ULXs.
Misra et al. (2020) modeled detailed evolution of low- and intermediate-mass
($\sim 092 - 8\,M_\odot$) X-ray binaries by using the MESA code and
considering beaming effects, tides, general angular momentum losses,
and a detailed/self-consistent calculation of the mass transfer
rate. They discovered that some observational properties of
pulsating ULXs can be reproduced by intermediate mass X-ray binaries
if non-conservative mass transfer and geometrical beaming are
adopted. They also studied detectable NS-white dwarf (WD) systems
that evolved from ULXs with intermediate-mass donors. Before this,
Marchant et al. (2017) and Finke \& Razzaque (2017) explored that
ULXs can be progenitors of compact objects as the gravitational wave
(GW) sources.

In this work, we studied formation and evolution of ULX pulsar
binaries by considering a wide range of physical information in the
binary evolution. We presented evolutions of binary systems to
pulsar ULXs with Be stars, helium stars and low, intermediate and
massive stars (including non-Be main-sequence and (sub)giant-branch
stars). We also investigated the final outcomes of ULXs and their
contribution to the compact objects (NS-NS and NS-WD systems) as the
gravitational sources. In \S 2, we describe our method to treat the
binary physical processes in the simulation. The calculated
properties and birth rates are given in \S 3. The ULX contributions
of NS-NS and NS-WD systems to LISA detection are also discussed in
\S 3. The discussion and conclusion are in \S 4.

\section{Binary Evolution Model and Calculations}
\label{sec:model}

\subsection{Simulation of Binaries}

We simulate the binary evolutionary processes which start from a
primordial system of two zero-age main-sequence (ZAMS) stars, and
main processes include mass transfer, stellar winds, tidal
interaction, angular momentum evolution, supernova explosions and
natal kicks, common envelope (CE) evolution, magnetic braking, GW
radiation, merger, etc.

We run a large number ($3\times10^7$ ) of binary evolution
calculations by applying the updated $BSE$ population synthesis code
based on Hurley et al. (2000, 2002). For the SN remnant calculation,
we adopted the rapid remnant-mass model of Fryer et al.(2012) in the
subroutine $hrdiag$ of the updated code, and note that a typo in
Fryer et al. (2012) has been corrected in our code (corrected as
$a_1 = 0.25 - \frac{1.275}{(M_1 - M_{\rm proto})}$, also in the
calculations of Ablimit \& Maeda (2018)). In the subroutine $evolv2$
of this updated $BSE$ code, we employ a mass transfer model related
to the rotation of the accretor (Ablimit et al. 2016), because
Ablimit \& Maeda (2018) found that profiles of formed compact
objects with this model are consistent with observed properties of
the gravitational sources.  Instead of using empirical results for
the critical mass ratio of Hurley et al. (2002), the numerically
calculated one based on the rotation-dependent mass transfer model
by considering both the response of the secondary to mass accretion
and the effect of possible mass loss (see Ablimit et al. (2016)) is
used. In this non-conservative and rotation-dependent mass transfer
model, the rotation of the secondary star (the star with a lower
mass is defined as the secondary) affects its accretion rate by a
factor of (1 - $\omega/\omega_{\rm cr}$), where $\omega$ is the
angular velocity of the secondary star and $\omega_{\rm cr}$ is its
critical value (Petrovic et al. 2005; de Mink et al. 2009). The
secondary star can spin faster to close its its critical value if
the star accretes a small amount of mass (Packet 1981), and this
makes the mass transfer efficiency as low as $<0.2$. Thus, the
maximal initial mass ratio of the primary to the secondary stars
in primordial binaries for avoiding the contact phase can
reach $\sim$5-6, and a larger number of the primordial binaries can
experience stable mass transfer phases until the primary's envelope
is completely exhausted (e.g., de Mink et al. 2013). If the
mass transfer is unstable (the primary is a evolved star), the
binaries are assumed to be a contact ones and enter a CE phase. The
two stars can merge into a single star with the other case.

For the CE evolution of a binary, the standard energy conservation
equation (Webbink 1984) is adopted in the subroutine $comenv$ of the
code,
\begin{equation}
E_{\rm{bind}} = {\alpha_{\rm CE}} \Delta E_{\rm orb} \ ,
\end{equation}
where $E_{\rm{bind}}$, $\alpha_{\rm CE}$, and $\Delta E_{\rm orb}$ are the binding energy of the envelope,
the efficiency parameter and the change in
the orbital energy during the CE phase, respectively. $\alpha_{\rm CE}=1.0$ is taken in this work.
The binding energy of the envelope
is expressed by the following:
\begin{equation}
E_{\rm{bind}} = - \frac{GM_1 M_{\rm{en}}}{\lambda {R}_1},
\end{equation}
where $M_1$, $M_{\rm{en}}$ and ${R}_1$ are the total mass, envelope
mass and radius of the primary star, respectively. The binding
energy is related to the stars' mass and radius (with different
evolutionary stages and metallicities). In the subroutine $comenv$
of the updated code, we use the results of Wang et al. (2016) for
the binding energy parameter $\lambda$ of the donor envelope. Very
recent CE study of Klencki et al. (2020) suggest that merger rates
from previous works could be severely overestimated, especially at
low metallicity due to the selected model of the binding energy
parameter. Their results and discussions support the results of Wang
et al. (2016). The orbital energy of the embedded binary is used to
expel the envelope. If binaries cannot survive from the CE, they
will merge into single stars,  and cannot contribute to the ULX
population. Binaries with stable mass transfer can avoid the CE, and
the envelope of the donor stars will be reduced through Roche lobe
overflow (RLOF). In the case of stable mass transfer, we simulate
the binary orbital evolution by assuming that the ejected matter
removes the specific orbital angular momentum of the accretor. For
the wind mass loss, the prescriptions of Vink et al. (2001) were
used for O and B stars in different stages (hot stars), and
$1.5\times10^{-4}\,\dot{M}_\sun\,\rm yr^{-1}$ (Vink \& de Koter
2002) is applied for luminous blue variable stars in the subroutine
$mlwind$ of the updated code (see the wind model 2 of Ablimit \&
Maeda (2018)). The prescription of the stellar wind mass losses in
Hurley et al. (2000) are adopted for other type stars, such as
Wolf-Rayet stars, cool red-giant-branch stars, etc.

The initial mass function of Kroupa et al. (1993) is utilized for
the primary star in the Monte-Carlo sample generating subroutine
$sample$ of the code,
\begin{equation}
f(M_1) = \left\{ \begin{array}{ll}
0 & \textrm{${M_1/M_\odot} < 0.1$}\\
0.29056{(M_1/M_\odot)}^{-1.3} & \textrm{$0.1\leq {M_1/M_\odot} < 0.5$}\\
0.1557{(M_1/M_\odot)}^{-2.2} & \textrm{$0.5\leq {M_1/M_\odot} < 1.0$}\\
0.1557{(M_1/M_\odot)}^{-\alpha} & \textrm{$1.0\leq {M_1/M_\odot} \leq 150$},
\end{array} \right.
\end{equation}
with $\alpha = 2.7$ in this study. The secondary mass distribution follows the distribution of the initial mass ratio,
\begin{equation}
n(q) = \left\{ \begin{array}{ll}
0 & \textrm{$q>1$}\\
\mu q^{\nu} & \textrm{$0\leq q < 1$},
\end{array} \right.
\end{equation}
where $q=M_2/M_1$ and $\mu$ is the normalization factor for the
assumed power law distribution with index $\nu$. We consider a flat
distribution ($\nu = 0$ and $n(q)=$constant) for the initial mass
ratio distribution (IMRD). The distribution of the initial orbital
separation, $a_{\rm i}$, is (Davis et al. 2008),
\begin{equation}
n(a_{\rm i}) = \left\{ \begin{array}{ll}
0 & \textrm{$a_{\rm i}/R_\odot < 3$ or $a_{\rm i}/R_\odot > 10^{6}$}\\
0.078636{(a_{\rm i}/R_\odot)}^{-1} & \textrm{$3\leq a_{\rm i}/R_\odot \leq 10^{6}$} \ .
\end{array} \right.
\end{equation}
The uniform (flat) initial eccentricity distribution is assumed in a range between 0 and 1.

Massive stars can evolve and become NSs through core-collapse SNe
(CCSNe) (a star with initial mass $\geq 10M_\odot$) or
electron-capture SNe (a star with initial mass between $8M_\odot
\geq$ - $\leq10M_\odot$) \footnote{Binary interaction could
change the masses of the stars, thus the initial mass limits for
leading CCSNe and ECSNe would be different compared to single stars.
Knowing core masses of stars in binaries would be more useful to
determine which of them go through ECSNe or CCSNe, and the relation
of initial masses and core masses has been studied extensively (see
recent works Sukhbold et al.(2016) and Woosely (2019) for more details).},
and both SNe channels are considered in this work (e.g., Fryer et
al. 2012). The SN explosion will impart a natal kick, and it causes
eccentric orbit or disruption of the binary system if it is a strong
kick. The kick velocities are assumed to obey Maxwellian
distributions with a dispersion of $\sigma= 40\,\rm\,km\,s^{-1}$
(Dessart et al. 2006) for NSs formed from electron-capture SNe
(ECSNe) and $\sigma = 265\,\rm\,km\,s^{-1}$ (Hobbs et al. 2005) for
NSs formed from CCSNe. In both channels, the mass of NSs is limited
between $\sim$1.0 and 3.0 $M_\odot$, and the majority of NSs in our
simulation results have masses in a range of $\sim$1.0 and 2.0
$M_\odot$.

The accretion-induced collapse of an ONe WD is suggested as one
important alternative formation way for NS binaries (e.g., Nomoto \&
Kondo 1991; Ablimit \& Li 2015; Zhu et al. 2015; Ablimit 2019; Wang
\& Liu 2020). In this work, we study the contribution of the AIC
channel to ULX pulsar binaries, and we employ the same method as
Ablimit (2019) for the accretion process (e.g., retention efficiency
of the transferred matter on the WD surface) of WD binaries.
Although it was pointed out that the magnetic field of the WD may
play a very important role in WD binary evolution (see Ablimit \&
Maeda 2019a, b; Ablimit 2019), we studied the non-magnetic case in
this work. The default values of other physical parameters and other
physical processes are the as same as described in Hurley et al.
(2000, 2002).

The main evolution routes from close ZAMS + ZAMS binaries to NS $+$
normal stars (including helium stars) binaries are very briefly
described as follows: Route 1: a massive ZAMS star ($\geq$ 10
$M_\odot$) binary with a relatively unit mass ratio or relatively
larger mass ratio starts their evolution, the more massive star
evolves and first fills it Roche lobe (RL), then a stable or
unstable RLOF mass transfer may occur. After the stable mass
transfer ends or surviving from the CE, the primary massive one
undergoes a CCSN and leaves a newborn NS, and forms an NS $+$ normal
star binary. Route 2: a massive ZAMS star (between 8 and 10
$M_\odot$) binary starts its evolution, and in other ways is similar
to above evolution. There is an ECSN of the primary and forms an NS,
then an NS $+$ normal star binary begins its evolution. Route 3: an
intermediate-mass ZAMS star (between 2 and 8 $M_\odot$) binary with
a relatively bigger mass ratio. The more massive star evolves more
and first fills its RL, then an unstable mass transfer may cause the
CE phase. If the binary survives from the CE,  the primary becomes a
WD, then forms a WD $+$ normal star binary. Later, the secondary
fills its RL and stars the stable mass transfer. The WD accretes
matter and grows in mass, and finally the WD collapses to form an NS
when the WD's mass reaches the Chandrasekhar limit mass (1.44
$M_\odot$ is adopted in this work), then an NS $+$ normal star
binary starts its evolution. In the section that showcases results,
we provide the initial parameter distributions for these three
routes.


\subsection{Calculations for NS ULXs}

The NS can be surrounded by matter which comes from the donor star,
and an accretion disk can be formed around the NS.
 Matter moves from outside to inside of the disk,
and the angular momentum transfers from inside to outside. The
accretion model is crucial for the evolution of the system including
the NS, and for observational signatures. If the mass from the donor
transfers to the NS at a sub-Eddington rate, the formed accretion
disk will be a thin disk, and it will become a thick disk when the
mass transfer has a super-Eddington rate (Shakura \& Sunyaev 1973).
This accretion model is used for both RLOF and
wind\footnote{For the wind accretion, we have adopted the
Bondi \& Hoyle (1944) accretion mechanism. The compact accretor
captures a fraction of the mass lost from the donor by stellar wind.
The prescription of Hurley et al. (2002) is applied for the wind
accretion factor which determines the mean accretion rate into the
disc around the compact accretor. The rest of the calculation is
treated in the same way as for RLOF accretion.} mass accretion.

At the super-Eddington rate, the accretion luminosity is Eddington
limited. The accretion luminosity can reach the Eddington luminosity
at the spherization radius which is,
\begin{equation}
R_{\rm sph}=\dot{M}_{\rm d}\frac{GM_{\rm NS}}{L_{\rm Edd}},
\end{equation}
where $\dot{M}_{\rm d}$ and $M_{\rm NS}$ are the mass transfer rate
from the donor star and mass of the NS, respectively. $G$ is the
gravitational constant. Outside $R_{\rm sph}$ the disk emits X-rays
with luminosity the Eddington $L_{\rm Edd}$,
\begin{equation}
L_{\rm Edd} = \eta \dot{M}_{\rm Edd} c^2 = 2.6\times10^{38}(\frac{1}{1+X})(\frac{M_{\rm NS}}{M_\odot})\,{\rm erg}\,{\rm s}^{-1},
\end{equation}
where $c$ is the speed of light. The radiative
efficiency\footnote{Note that the radiative efficiency would
vary according to the NS mass.} of accretion onto the NS is $\eta
=0.1$ in this work. The Eddington limit for the mass accretion rate
is $\dot{M}_{\rm Edd} = L_{\rm Edd}/(\eta c^2)$ for an NS with a
mass of $M_{\rm NS}$. $X$ is the hydrogen mass fraction in the donor
envelope, and it is taken as 0.7 for H-rich donor stars and 0 for
H-deficient donor stars in this work. Inside of  the spherization
radius ($R_{\rm sph}$), the disc is dominated by radiation pressure.
Strong outflows begins and the accretion rate decreases linearly
with radius. The mass accretion rate within $R_{\rm sph}$ at each
point in the local disk follows the relation,


\begin{equation}
\dot{M}^L_{\rm Edd}(R) = \dot{M}_{\rm d}\frac{R}{R_{\rm sph}}.
\end{equation}

Because of the geometric collimation, one can see the source in
directions within one of the cones, with an apparent (isotropic)
X-ray luminosity. Given the mass transfer rate, we calculate the X-ray
luminosity according to King (2008, 2009),

\begin{equation}
L_{\rm X} = \frac{L_{\rm Edd}}{b}  (1 + {\rm ln}\dot{m}),
\end{equation}
where the beaming factor $b$ is,
\begin{equation}
b = \left\{ \begin{array}{ll}
\frac{73}{{\dot{m}^2}} & \textrm{if $\dot{m}>8.5$}\\
1 & \textrm{otherwise} \ .
\end{array} \right.
\end{equation}
where $\dot{m} = \dot{M}_{\rm d}/\dot{M}_{\rm Edd}$. In order to
avoid the unphysical very high values when we approximate the
beaming factor, the calculated accretion (apparent) luminosity is
limited by $10^{39} \leq L_{\rm X} \leq 10^{44}\,\rm erg\,s^{-1}$.

In the accreting NS binaries, NSs can be easily spun up to be
(millisecond) pulsars during the evolution processes by mass
accretion, and the spin rates of the recycled pulsars would be
slower if the companion stars are more massive and/or giant stars
due to the shorter timescale of the mass-transfer (X-ray) phase of
massive and/or giant stars (e.g., Tauris, Langer \& Kramer 2012).
For a slow or non-rotator newborn NS, the amount of accreted mass
needed to spin up as a pulsar has been studied and discussed by
Tauris, Langer \& Kramer (2012). An NS with a typical mass of 1.4
$M_\odot$ can spin up to 50 milliseconds from rest if it accretes
only a small amount of mass ($10^{-3}\,M_\odot$) (see the equations
and discussions of Tauris, Langer \& Kramer (2012) for more
details). Observed spin periods of ULXs pulsars (see Table 1) are at
the level of level which naturally can be achieved by the
mass-transfer (X-ray) phases of ULXs. In our calculation, most of
the NSs in ULXs have pulse periods distributed between $\sim$ 0.3
and 10 seconds. Thus, we assume all NSs in ULX binaries can be
pulsars, and we do not trace after the spin period evolution of the
NS in this study. We investigate the formation of ULX binaries, and
try to explain the observed orbital periods, apparent luminosities
and indicated masses of known ULX pulsar binaries.

\section{Results}

We simulate evolutions of MS-MS binaries which evolve toward ULX
pulsar binaries with different types of companion stars, and we also
explore the final outcomes of ULX pulsar binaries. In this work, the
type of non-degenerate companion stars (its mass is $M_{\rm d}$) are defined as follows: the
stars at the central hydrogen burning stage are assumed as the MS
stars; cases that finished all central hydrogen burning are
considered as evolved stars; The evolved normal stars include
sub-giant stars (hydrogen-rich Hertzsprung gap stars), and giant
branch stars (red giant stars, core helium burning stars, etc.);
stars with masses $< 2$ $M_\odot$ are low-mass stars; stars with
masses $2\leq M_{\rm d} < 8\,M_\odot$ are intermediate-mass stars; stars
with masses $\geq 8$ $M_\odot$ are massive stars; Rapidly
rotating B-type stars are categorized as Be stars, and rotation
speed of Be stars generally reaches nearly their Keplerian limits.
If spin angular velocities of stars are close to their critical spin
angular velocities ($\frac{\omega}{\omega_{\rm cr}} > 0.85$), they
are selected as Be stars;  Note that Be stars and other  slow
rotator stars with $\frac{\omega}{\omega_{\rm cr}} \leq 0.85$ (MS
and evolved stars) are seen as normal stars; Stars are categorized
as Helium stars if their hydrogen-rich envelope are stripped away by
the binary interactions, and He MS, He-rich Hertzsprung gap and He
giant stars are considered as He stars in this work.

\subsection{Mass, orbital period and luminosity profiles of ULX pulsar binaries}

In this subsection, we show calculated orbital periods ($P_{\rm orb}$), masses and
accretion luminosities of ULX pulsar binaries, and compare our
results with those of observed ULX pulsar binaries. In Figure 1, the
red triangles represent the data of observed ULX pulsar sources, and
the other colors and symbols show the simulated ULXs with different
type of companions. Because the observational mass predictions have
uncertain values (see Table 1, and see Section 1 for discussion), we
take values within the observational indications for companions'
masses and a few uncertain orbital periods of seven observed ULXs,
such as $M_{\rm d}=10\,M_\odot$ and $P_{\rm orb}=380$ days for NGC
300 ULX-1; $M_{\rm d}=6.0\,M_\odot$ and $P_{\rm orb}=6$ days for the
source NGC 1313 X-2; $M_{\rm d}=5.2\,M_\odot$ for the source M82
X-2; $M_{\rm d}=20\,M_\odot$ for the source NGC 7793 P13; $M_{\rm
d}=4.0\,M_\odot$ for the source NGC 5907 ULX-1; $M_{\rm
d}=3.7\,M_\odot$ for the source SMC X-3; $M_{\rm d}=6.0\,M_\odot$
for the source M51 ULX-1. The donor mass and orbital period
distributions from our calculations show that Be stars, massive
stars and intermediate mass stars are more likely companion stars to
ULXs as their values match the observationally indicated values of
seven ULX pulsar binaries. The luminosity distributions in Figure 2
demonstrate that NS binaries with different types stars can be
higher than $L_{\rm X} \geq 10^{39}\,erg\,s^{-1}$, and at least 80\%
of them have luminosities distributed in $10^{39} \leq L_{\rm X}
\leq 10^{41}\,\rm erg\,s^{-1}$. The apparent luminosities of all
nine known pulsar ULXs (Table 1) can be reproduced by the
simulation.

Although binaries with low-mass stars can be companions to ULX
pulsar binaries, the derived masses and most orbital periods of
low-mass stars are not consistent with currently available
observational mass indications of seven sources. He stars can also
contribute to ULX pulsar systems, however, the mass of He stars in
ULX systems is mainly distributed between 0.6 and 2.8 $M_\odot$
which are lower than observationally predicted ones (wider than the
mass range from Wiktorowicz et al.(2019)). The systems with a
massive helium star are hardly produced because of a combination of
relatively low birthrates, short mass transfer durations, and low
detection possibilities (due to the beaming effect). ULXs with He
stars have age from 18 Myr to 210 Myr. The orbital period
distribution of ULX pulsar binaries with He stars has a peak at 0.1
day, and the longest one is up to a few hundred days.
The systems containing NSs and He stars are also promising multi-messenger
sources for the upcoming electromagnetic and GW facilities (e.g.,
G$\ddot{\rm o}$tberg et al. 2020), but this is not the focus of this
work. We discuss normal star companions in more details below.

Figure 3 shows the initial masses and initial orbital period
distributions of progenitors of ULX pulsar binaries with normal star
donors. Distributions are basically consistent with results of
previous BPS studies (e.g., Fragos et al.2015). 91.5\% of NSs in ULX
systems with normal star companions are born through CCSNe, 4.5\% of
NSs in ULXs are formed through the AIC of massive WDs, and 4.0\% of
NSs in ULX binaries are produced through electron-capture SNe (see
the above section for more discussions of the three routes). The
main properties of ULX pulsar binaries are shown in Figure 4. Donor
mass distributions of pulsar ULXs show that 8.9\% of pulsar ULXs
have low mass companions while 91.1\% of them have intermediate or
massive star companions. The majority of NS ULXs were believed to
contain red giant companions with typical masses around 1 $M_\odot$
by Wiktorowicz et al.(2017). However, masses indicated by recent
observations for seven known ULX pulsar systems (see Table 1) imply
that Be stars, massive stars and/or intermediate-mass stars could be
companion stars in these pulsar binaries, which are consistent with
what our BPS simulations find. Most ULXs in this work have orbital
periods between $\sim$ 1 and 10 days (Figure 4), and this also
supports the statements given by BPS results of Fragos et al.(2015)
and detailed binary evolution results of Misra et al.(2020). ULX
binaries with Be stars have age between 12 Myr and 217 Myr
in our calculation, but the ages of other normal stars are
mainly distributed between 18 Myr and 1 Gyr (up to several Gyr).

Formation of ULX pulsar binaries containing Be star donors is more
common in our simulation. Their orbital periods are distributed from
$>1$ day to $\sim$1000 days, and Be star masses have a range of 2.5
$M_\odot$ - 30 $M_\odot$. These properties cover the observationally
indicated masses and orbital periods of all seven ULX pulsar sources
displayed in Table 1 and Figure 1 (at least covering five possible
candidates known as Be ULXs which exhibit transient phases of X-ray
emission, such as NGC 300 ULX1 (Binder et al. 2016) and SMC X-3
(Townsend et al. 2017)). One of the main features of a Be star is
fast rotation which is considered in this work by including the mass
transfer and tidal effect, and binary interactions like the mass
transfer and tidal effect have been suggested as possible mechanisms
for fast rotation of the star and formation of ULXs (e.g., Ablimit
\& L$\ddot{\rm u}$ 2013). An excretion disk would be formed
around Be stars due to the fast rotational velocities of Be stars.
NSs in most NS-Be star binaries (Reig 2011) accrete matter when they
pass through or are close to the Be star's circumstellar disc.
Theoretical models usually assume that a Be star is just a rapid
rotator, and it is very difficult to include the Be disk in the 1D
modeling. There are some important details in the formation of
circumstellar discs around Be stars and origin of different types of
outbursts in galactic and extragalactic Be stars, which need further
investigations (Lee, Osaki \& Saio 1991; Negueruela et al. 2001;
Negueruela \& Okazaki 2001; Martin et al. 2014). These physical
processes are out of the scope of this work.

\subsection{Rates of the ULX pulsar binary}

The birth rate ($R_{\rm B}$) of a type of binary can be calculated
by the following equation,
\begin{equation}
R_{\rm B} = f_{\rm bin}\times\eta_{\rm SFR}\times\frac{\rm N}{M_{\rm total}}  ,
\end{equation}
where N is the total number of a specific binary types, and $M_{\rm
total}$ is the total mass of all stellar systems. $\eta_{\rm SFR}$
is the star formation rate. $f_{\rm bin}$ is a fraction of binaries
in all stellar systems, which is $f_{\rm bin} = \frac{N_{\rm
bin}}{N_{\rm bin}+N_{\rm single}}$.

In Table 2, we show the formation rates of ULX pulsar binaries per
$10^6 M_\odot$ of created stars in the simulation with different
donors. If we assume a constant star formation rate of $\eta_{\rm
SFR}=3M_\odot\,\rm yr^{-1}$ and binarity of $f_{\rm bin} = 0.7$, we
obtain that the birth rates of ULX pulsar binaries with normal stars
and He stars are $\sim 1.83\times10^{-4}\,\rm yr^{-1}$ and $\sim
3.36\times10^{-5}\,\rm yr^{-1}$ at the solar metallicity case,
respectively. We also calculated the formation rates for the
population II stars (low metallicity case), and the rates become
higher with the case of $Z=0.001$. Be stars are main donors for ULXs
in our calculation, and the evolved stars (HG or/and giant branch
stars) are much less than MS donors. Only among low-mass donors, the
evolved stars are comparable with low mass MS stars, while it is
just $\sim$4.0\% and 3.7\% among intermediate and high mass donors
(MS donors are significantly more), respectively. Wiktorowicz et al.
(2017, 2019) only considered RLOF mass transfer case in their study,
but they did not include the wind-fed mass transfer (which plays an
important role in massive star binary evolution), thus it may cause
an underestimation of the fraction of high mass donors in ULXs
(about 1\%). A recent ULX observational result of L$\acute{\rm
o}$pez et al. (2020) indicates that the fraction of giant massive
stars could be around $4\pm2$\% (note that they did not categorize
the accretor of ULXs in their observational result). This is higher
than what Wiktorowicz et al. (2017) find, but it is supported by our
results. We also show the rate evolutions of ULXs with the constant
star formation model and a single burst star formation model in
Figure 5. The large number of ULX pulsar binaries at solar
metallicity needs a shorter timescale than that of the low
metallicity case, and they all have a peak in the first $<$ 100 Myr
(a significant number of NS ULXs have their ULX phases during the
post MS expansion of their donors at low metallicity). Our results
suggest the solar metallicity condition for producing current
observational results of pulsar ULXs.

\subsection{Evolution of ULX pulsar binaries to Pulsar-NS and Pulsar-WD systems for the LISA detector}

Except for ULXs with very low mass donors, ULXs with
intermediate-mass donors and He stars naturally evolve to form
pulsar - WD binaries in the Hubble time, and they have a large
fraction of whole ULXs. A pulsar-NS system also can be expected from
ULXs with massive stars. From the current understanding of binary
evolution, they also may evolve into the CE phase during the
evolution of ULX pulsar binaries, and they will form double compact
objects in close orbits if they survive from the CE, otherwise they
will merge inside the CE and become one object. The connections
between ULXs and double compact star objects (DCOs) can potentially
be important to understand the origin of merging DCOs detected by
LIGO. We analyze our BPS results with the solar metallicity case and
try to find those connections.

In this work, we find that pulsar ULXs would become pulsar-WD and
pulsar-NS systems (no other DCOs). After a pulsar-WD binary formed
in a very short orbit, the WD can fill its Roche lobe and the mass
transfer may proceed. The merger or an ultra-compact X-ray binary
(UCXB; e.g., Heinke et al. 2013; L$\ddot{\rm u}$ et al. 2017) can be
expected depending on the stability of the RLOF mass transfer.
Because we derived similar merger rates with different models, we
assume the final stage of a pulsar-WD system will be a merger
through emitting GW signals.

The orbital motion of a pulsar-WD or pulsar-NS binary system can
lead to GW emission and spiral-in of the system. The continuous
energy loss through GW emission (according to general relativity)
will merge the two objects at the final stage of the inspiraling
process. The timescale (Peters \& Mathews 1963; Peters 1964; Lorimer
2008) of the merging process is expressed as,

\begin{equation}
t = 9.88\times10^{6}(\frac{P_{\rm orb}}{ 1 \,\rm hr})^{8/3} (\frac{\mu}{1\,M_\odot})^{-1} (\frac{M}{1\,M_\odot})^{-2/3}\, {\rm yr},
\end{equation}
where $\mu = M_1M_2/(M_1 + M_2)$ and $M = M_1 + M_2$ are the reduced
mass and total mass of the system ($M_1$ and $M_2$ are the masses of
two objects in a binary system), respectively.

The continuous GW emission takes away orbital angular momentum and
shrinks the orbit, and the GW strain will increase with time during
the merging process. If the distance of the binary system with
respect to us is $d$, then the strain amplitude of the GW can be
expressed as (Peters \& Mathews 1963)

\begin{equation}
h = 5.1\times10^{-23}(\frac{P_{\rm orb}}{ 1 \,\rm hr})^{-2/3} (\frac{M_{\rm ch}}{1\,M_\odot})^{5/3} (\frac{d}{10\,{\rm kpc}})^{-1},
\end{equation}
where $M_{\rm ch} = (M_1M_2)^{3/5}/(M_1 + M_2)^{1/5}$ is the chirp mass. The GW frequency $f$ is twice the orbital motion
frequency.

From the results of the BPS simulation, we derive that rates of
NS-WD and NS-NS systems that can merge in a Hubble time are
$1.28\times10^{-4}\,{M_\odot}^{-1}$ and
$4.34\times10^{-5}\,{M_\odot}^{-1}$, respectively. If a constant
star formation rate of $\eta_{\rm SFR}=3M_\odot\,\rm yr^{-1}$ and
binarity of $f_{\rm bin} = 0.7$ are adopted, the merger rate of
NS-WD and NS-NS systems that can merge in a Hubble time are
$2.69\times10^{-4}\,{\rm yr}^{-1}$ ($\sim$2690 ${\rm Gpc}^{-3}{\rm
yr}^{-1}$) and $0.91\times10^{-4}\,{\rm yr}^{-1}$ ($\sim$910 ${\rm
Gpc}^{-3}{\rm yr}^{-1}$) in Milky Way-like galaxies, respectively.
The rate of double NS merger in this work is consistent with the
observationally inferred rate by LIGO (250 - 2810 ${\rm
Gpc}^{-3}{\rm yr}^{-1}$) (Abbott et al. 2020). Among them, at least
$\sim$19.2\% of NS-WD merger signal and $\sim$7.8\% NS-NS merger
signal can be detected by LISA. $\sim$41\% of all the NS-WD
population is pulsar-WD systems that evolved from ULX pulsar
binaries, while $\sim$4.2\% of those NS-NS systems is pulsar-NS
systems formed from ULX pulsar binaries (Figure 6). $\sim$13\% of
ULXs with He stars and $\sim$87\% of ULXs with normal stars
contribute to the pulsar-WD GW sources, and all pulsar-NSs evolved
from ULXs with normal stars. From the whole NS-WD and NS-NS
populations that merge in the Hubble time, we find that
contributions of pulsar-WD and pulsar-NS systems from ULX pulsar
channels for LISA are $\sim$10.5\% and $\sim$1.0\% (as shown in
Figure 6), respectively. Note that the finite size of the Monte
Carlo simulation gives a statistical fluctuation. With different BPS
models (Ablimit et al. 2016; Ablimit \& Maeda 2018), the relative
statistical error is estimated to be $\sim$5\%--35\%.


\section{Discussion and Conclusion}

The observational results of ULXs are increasing, and the properties
of known pulsar ULXs need further investigations. We simulated a
large number of binary evolutions, and analyzed physical processes.
In the primordial binary evolution, we adopted a non-conservative
and rotation-dependent mass transfer model (numerically calculated
critical mass ratio) which allows most primordial systems to have
stable mass transfer, and this makes more NSs have relatively higher
mass donors (most of them are intermediate mass donors). Thus, we
have a few percent higher high mass donor ULXs compared to previous
works, and have significantly more intermediate mass ULXs as shown
in the results section. Formation of NS ULXs with normal stars in
this work is less dependent on the ejection of CE compared to
previous works. The mass transfer model and metallicity have
important roles in the formation and evolution of ULXs. The later
evolution of more NS + normal star binaries experience the CE phase,
and a more reliable model of binding energy for the CE evolution is
adopted in the work. The main results can be concluded as follows:

\begin{enumerate}
\item Our BPS simulations tend to have more pulsar ULXs with Be stars and intermediate mass donors.
The observationally indicated properties of seven known pulsar ULXs can be fully reproduced by the combined results of Be and intermediate mass ULXs.
Few known ULXs also can be formed with high mass donors.
NS ULXs with He star or low mass donors are not consistent with current observations of seven pulsar ULXs, although they have obvious
contributions to NS ULXs in the simulation results. The rate (number) of pulsar ULXs can be higher with the population II ($Z=0.001$) case
than that of the population I ($Z=0.02$) case.

\item The AIC route has 4.5\% contribution to the NS ULXs. 4.0\% of NSs in ULXs are formed through ECSNe, while 91.5\% of NSs in ULXs are born with CCSNe.
The fraction of pulsar ULXs with MS donors is significantly larger
than that of ULXs with evolved donors. The fractions of evolved
hydrogen-rich low, intermediate and high mass donors in the whole
normal star donors are 3.3\%, 3.5\% and 3.2\%, respectively.

\item We do not find pulsar - BH systems evolved from pulsar ULXs who can merge in the Hubble time. We discover that
pulsar-NS and pulsar-WD systems from ULXs have $\sim$4.2\% and 41\%
contributions to the whole NS-NS and NS-WD systems which merge in a
Hubble time. Pulsar-NS systems evolved from ULXs with normal stars,
while $\sim$13\% ULXs with He stars and $\sim$87\% ULXs with normal
stars evolved to form pulsar-WD binaries which can merge in the age
of the Universe. Among the all detectable NS-NS and NS-WD systems
for the LISA detector, $\sim$11\% and 40\% are pulsar-NS and
pulsar-WD systems that evolved from ULXs. The Monte Carlo simulation
uncertainties with different BSP models should be considered when
dealing with these contributions.

\end{enumerate}

\begin{acknowledgements}

This work was supported by the National Natural Science Foundation of
 China, project No. 11863005.
The LAMOST FELLOWSHIP is supported by Special Funding for Advanced Users, budgeted and
administrated by Center for Astronomical Mega-Science, Chinese Academy of Sciences.

\end{acknowledgements}


\begin{center}
REFERENCES
\end{center}

Abbott, B. P., Abbott, R., Abbott, T. D., LIGO collabortaion, et al., 2020, ApJL, 892, L3

Ablimit, I. \& Li, X.-D., 2015, ApJ, 800, 98

Ablimit, I. Maeda, K. \& Li, X.-D., 2016, ApJ, 826, 53

Ablimit, I. \& Maeda, K., 2018, ApJ, 866, 151

Ablimit, I. \&  L$\ddot{\rm u}$, GuoLiang, 2013, SCPMA, 56, 663

Ablimit, I. \& Maeda, K., 2019a, ApJ, 871, 31

Ablimit, I. \& Maeda, K., 2019b, ApJ, 885, 99

Ablimit, I., 2019, ApJ, 881, 72

Bachetti, M., et al., 2014, Nature, 514, 202

Binder, B., Williams, B. F., Kong, A. K. H., 2016, MNRAS, 457, 1636

Bondi, H. \& Hoyle, F., 1944, MNRAS, 104, 273

Brightman, M., et al., 2018, NatAs, 2, 312

Brightman, M., Balokovic, M., Koss, M., et al., 2018b, ApJ, 867, 110

Brightman, M., Harrison, F. A., Bachetti, M., et al., 2019,ApJ, 873, 115

Colbert, E. J. M., Mushotzky, R. F., 1999, ApJ, 519, 89

Carpano, S., Haberl, F., Maitra, C., et al., 2018, MNRAS, 476, L45

Chandra, A. D., Roy, J., Agrawal, P. C. \& Choudhury, M., 2020, MNRAS, 495, 2664

Dall'Osso S., Perna R., Stella L., 2015, MNRAS, 449, 2144

Davis, P. J., Kolb, U., Willems, B., Gnsicke, B. T., 2008, MNRAS,389, 1563

de Mink, S. E., Langer, N., Izzard, R. G., Sana, H., \& de Koter, A. 2013, ApJ,
764, 166

de Mink, S. E., Pols, O. R., Langer, N., \& Izzard, R. G. 2009, A\&A, 507, L1

Dessart, L., Burrows, A., Ott, C. D., et al. 2006, ApJ, 644, 1063

Doroshenko V., Tsygankov S., Santangelo A., 2018, A\&A, 613, A19

Finke, J. D.,  Razzaque, S., 2017, MNRAS, 472, 3683

Fabbiano, G., 1989, ARA\&A, 27, 87

Feng, H., Soria, R., 2011, New Astronomy Reviews, 55, 166

F$\ddot{\rm u}$rst, F., Walton, D. J., Harrison, F. A., et al., 2016, ApJ, 831, L14

F$\ddot{\rm u}$rst, F., Walton, D. J., Heida, M. et al., 2018, A\&A, 616, A186

Fragos, T., Linden, T., Kalogera, V., \& Sklias, P. 2015, ApJ, 802, L5

Fryer, C. L., Belczynski, K., Wiktorowicz, G., et al. 2012, ApJ, 749, 91

G$\ddot{\rm o}$tberg, Y.,  Korol, V., Lamberts, A. et al., 2020, arXiv:2006.07382

Heinke, C. O., Ivanova, N., Engel, M. C., et al. 2013, ApJ, 768, 184

Heida, M., Harrison, F. A., Brightman, M., F$\ddot{\rm u}$rst F., SternD., Walton D. J., 2019, ApJ, 871, 231

Heida, M., et al., 2019, ApJL, 883, L34

Herold, H., 1979, PhRvD, 19, 2868

Hobbs, G., Lorimer, D. R., Lyne, A. G., \& Kramer, M. 2005, MNRAS, 360, 974

Hurley, J. R., Pols, O. R., Tout, C. A.,  2000, MNRAS, 315, 543

Hurley, J. R., Tout, C. A., Pols, O. R., 2002, MNRAS, 329, 897

Israel, G. L., Papitto, A., Esposito, P. , et al., 2017a, Science, 355, 817

Israel, G. L., Papitto, A., Esposito, P., et al., 2017, MNRAS, 466, L48


Kaaret, P., Feng, H., \& Roberts, T. P. 2017, ARA\&A, 55, 303

King, A. R., 2008, MNRAS, 385, L113

King, A. R. 2009, MNRAS, 393, L41

King, A., Lasota, J.-P., 2019, MNRAS, 485, 3588

King, A., Lasota, J.-P., 2020, arXiv arXiv:2003.14019

King, A. R., Davies, M. B., Ward, M. J., et al. 2001, ApJ, 552, L109

Klencki, J., Nelemans, G., Istrate, A.-G. \& Chruslinska, M., 2020, arXiv:2006.11286

Klu$\acute{\rm z}$niak, W., Lasota, J. P., 2015, MNRAS, 448, L43

Kroupa, P., Tout, C. A.,  Gilmore, G., 1993, MNRAS, 262, 545

Lee U., Osaki Y. \& Saio H., 1991, MNRAS, 250, 432

L$\acute{\rm o}$pez, K. M., Heida, M., Jonker, P. G., et al., 2020, arXiv: 2006.02795

Lorimer, D. R. 2008, LRR, 11, 8

L$\ddot{\rm u}$, GuoLiang, Zhu, C.-H., Wang, Z.-j., Iminniyaz, H., 2017, ApJ, 847, 62

Martin R. G., Nixon C., Armitage P. J., Lubow S. H. \& Price D. J., 2014, ApJ,
790, L34

Misra, D., Fragos, T., Tauris, T. M., Zapartas, E. \& Aguilera-Dena, D. R. 2020, arXiv::2004.01205

Mushtukov, A. A., Suleimanov, V. F., Tsygankov, S. S., et al. 2015, MNRAS, 454, 2539

Marchant, P., Langer, N., Podsiadlowski, P., et al. 2017, A\&A, 604, A55

Negueruela I., Okazaki A. T., 2001, A\&A, 369, 108

Negueruela I., Okazaki A. T., Fabregat J., Coe M. J., Munari U., Tomov T.,
2001, A\&A, 369, 117

Nomoto, K., Kondo, Y., 1991, ApJ , 367, L19

Shakura, N. I., Sunyaev, R. A., 1973, A\&A, 24, 337

Packet, W., 1981, A\&A, 102, 17

Paxton, B., Marchant, P., Schwab, J., et al., 2015, ApJS, 220, 15

Petrovic, J., Langer, N., \& van der Hucht, K. A., 2005, A\&A, 435, 1013

Peters, P. C. 1964, PhRv, 136, B1224

Peters, P. C., \& Mathews, J. 1963, PhRv, 131, 435

Poutanen, J., Lipunova, G., Fabrika, S., et al. 2007, MNRAS, 377, 1187



Rajoelimanana, A. F., Charles, P. A., Udalski A., 2011, The Astronomer's Telegram, 3154

Reig, P. 2011, Ap\&SS, 332, 1

Rodr$\acute{\rm i}$guez Castillo, G. A., et al., 2019, arXiv arXiv:1906.04791


Sathyaprakash, R., Roberts, T. P., Walton, D. J., et al., 2019, MNRAS, L104

Shao, Y. \& Li, X.-D., 2015, ApJ, 802, 131

Sukhbold, T., Ertl, T., Woosley, S. E., et al., 2016, ApJ, 821, 38

Tauris, T. M., Langer, N. \& Kramer, M. 2012, MNRAS, 425, 1601

Tsygankov, S. S., Doroshenko, V., Lutovinov, A. A., et al., 2017, A\&A 605, A39

Townsend, L. J., Kennea, J. A., Coe M. J., et al., 2017, MNRAS, 471, 3878


Vasilopoulos, G., Ray, P. S., Gendreau, K. C., et al., 2020, MNRAS, 494, 5350

Vink, J. S., de Koter, A., \& Lamers, H. J. G. L. M. 2001, A\&A, 369, 574

Vink, J. S., \& de Koter, A. 2002, A\&A, 393, 543


Wang, B., \& Liu, D., 2020, RAA, 20, 133

Wang, C., Jia, K., \& Li, X.-D. 2016, RAA, 16, 126

Webbink, R. F. 1984, ApJ, 277, 35

Wiktorowicz, G., Sobolewska M., Lasota J.-P. \& Belczynski, K. 2017, ApJ, 846, 17

Wiktorowicz, G., Lasota, J.-P., Middleton, M., \& Belczynski, K. 2019, ApJ, 875,53

Walton, D. J., Furst, F., Bachetti, M. et al., 2016, ApJ, 827, L13

Woosely, S. E. 2019, ApJ, 878, 49

Zhu, C.-H., L$\ddot{\rm u}$, G.-L., \& Wang, Z. 2015, MNRAS, 454, 1725


\begin{table}

\caption{Observed properties of ULX pulsars}

\begin{tabular}{lllll}
 \hline\hline
ULX pulsars & $\rm L_x$ ($\rm erg\,s^{-1}$) & $\rm P_s$ (s) & $\rm P_{\rm orb} (d)$ & $M_2\,(M_\odot)$  \\
\hline
NGC7793 P1$3^{1,2,3}$ & $5\times10^{39}$ & 0.42 & 64.0 & 18.0-23.0 \\
NGC5907 ULX-$1^{4,5,6}$ & $\sim 10^{41}$ & 1.137 & 5.3 & 2.0-6.0 \\
M82 X-$2^{7,8}$ & $1.8\times10^{40}$ & 1.37 & 2.52 & $\geq5.2$ \\
NGC1313 X-$2^{9}$ & $1.5\times10^{40}$ & 1.46 & $< 6$ & $\leq12$ \\
M51 ULX-$7^{10,11,12}$ & $7.1\times10^{39}$ & 2.8 & 2.0 & $\geq8$ \\
SMC X-$3^{13,14,15}$ & $2.5\times10^{39}$ & 7.77 & 44.92 & $\geq3.7$ \\
Swift J0243.6+612$4^{16,17}$ & $1.8\times10^{39}$ & 9.86 & 28.3 & -- \\
NGC300 ULX-$1^{18}$ & $4.7\times10^{39}$ & 31.6 & $>300$ & $\geq8-10$ \\
NGC 2403 UL$\rm X^{19}$ & $1.2\times10^{39}$ & $\sim 18$ & 60-100 (?) & --\\

\hline
\end{tabular}

1, F$\ddot{\rm u}$rst et al. (2016); 2, Israel et al. (2017b); 3, F$\ddot{\rm u}$rst et al. (2018);
 4, Walton et al.(2016); 5, Heida et al.(2019); 6, Israel et al. (2017a);
 7, Bachetti et al. (2014); 8, Brightman et al. (2019); 9, Sathyaprakash et al. (2019);
 10, Rodr$\acute{\rm i}$guez Castillo, et al. (2019); 11, Brightman et al .(2018b) ; 12, Vasilopoulos et al. (2020);
 13, Doroshenko, Tsygankov \& Santangelo.(2018); 14, Townsend et al .(2017); 15, Rajoelimanana et al .(2011);
 16, Townsend et al. (2017); 17, Doroshenko et al.(2018); 18, Carpano et al. (2018); 19, Tsygankov et al. (2017).
Note: There is new source recently found, which is RX J0209.6-7427 (see Chandra et al. 2020).
\end{table}

\clearpage

\begin{table}

\begin{center}
\caption{Formation rates of ULXs pulsar binaries (with different type of companions)
per $10^{6}\,{\rm M_\odot}$ of created stars for the simulation}

\begin{tabular}{l l l l }
 \hline\hline
Companion Type & Rate with $Z=0.02$ & Rate with $Z=0.001$ \\
\hline
 LM MS & 5.4 &  9.3  \\
 LM evolved & 3.3 &  5.2 \\
 IM MS & 22.9 &  28.0  \\
 IM evolved & 3.5 &  7.3   \\
 HM MS & 10.2 &  18.8  \\
 HM evolved & 3.2 &  4.2 \\
 Be star & 39.0 &  61.6  \\
 He star & 16.0 &  45.8    \\
\hline
\end{tabular}
\end{center}
\end{table}

\clearpage

\begin{figure}
\centering
\includegraphics[totalheight=4.3in,width=5.0in]{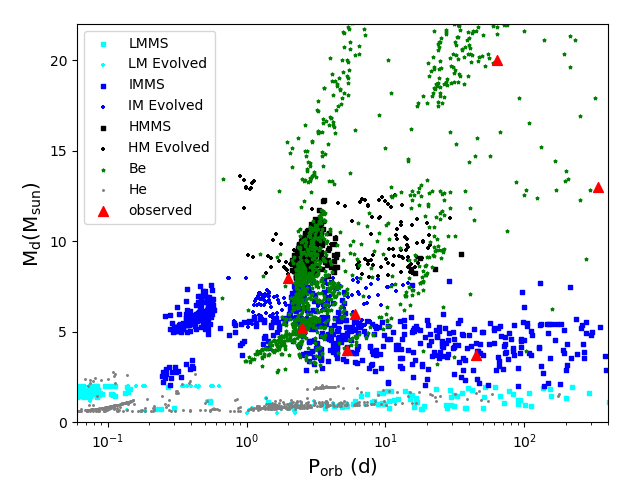}
\caption{The distributions of the orbital periods and companion stars' mass for ULX pulsar binaries.
Red triangles are observed sources (Table 1). ULX pulsar binaries from our results with Be star companions, He Star,
low-mass star (LM, including MS \& evolved), intermediate-mass star (IM, including MS \& evolved),
and high mass star (HM, including MS \& evolved) companions are shown, respectively.}
\label{fig:1}
\end{figure}

\clearpage

\begin{figure}
\centering
\includegraphics[totalheight=4.0in,width=4.5in]{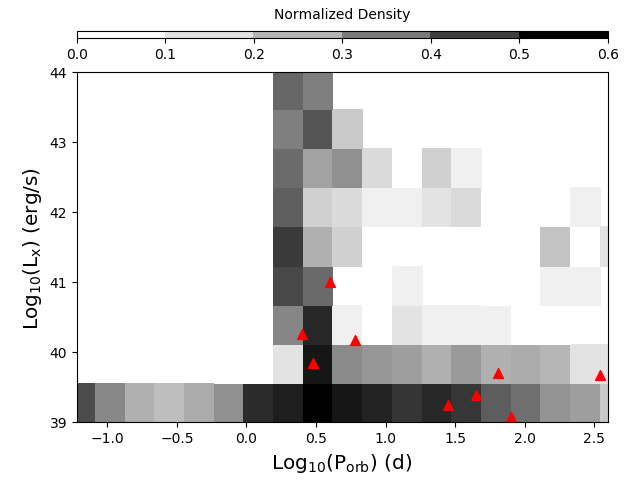}
\caption{The distributions of the orbital periods and accretion (apparent) luminosities for ULX pulsar binaries.
Red triangles are observed pulsar ULX sources.}
\label{fig:1}
\end{figure}

\clearpage

\begin{figure}
\centering
\includegraphics[totalheight=2.85in,width=3.8in]{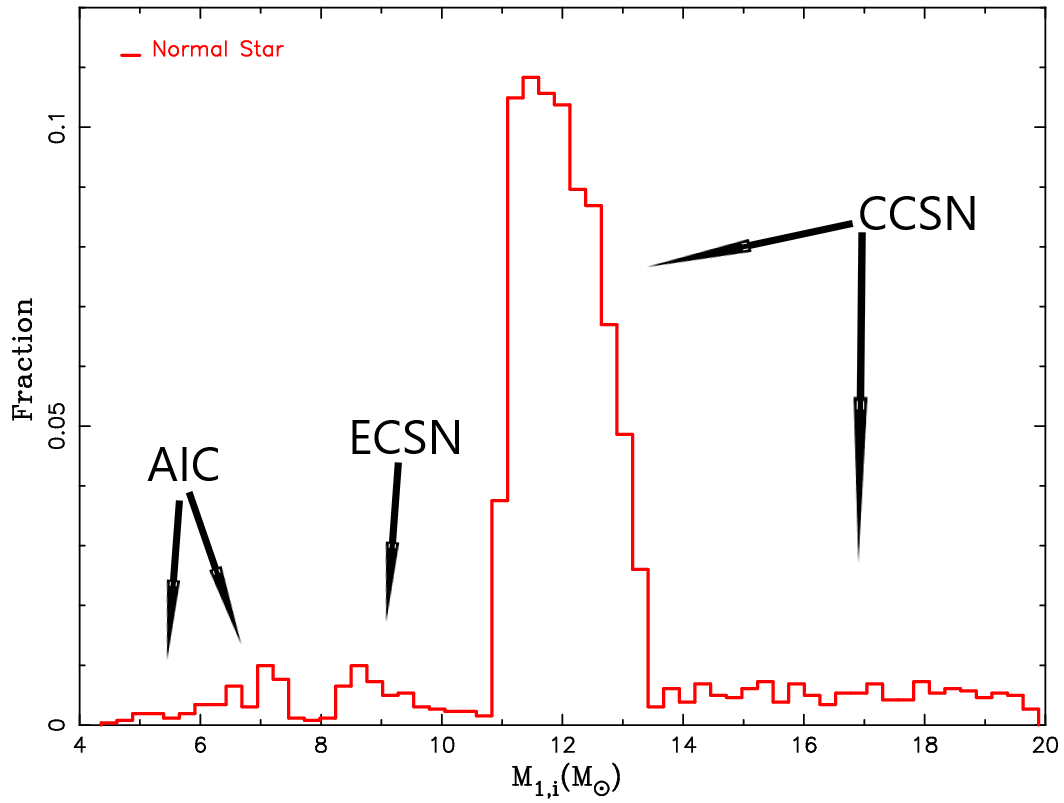}
\includegraphics[totalheight=2.85in,width=3.8in]{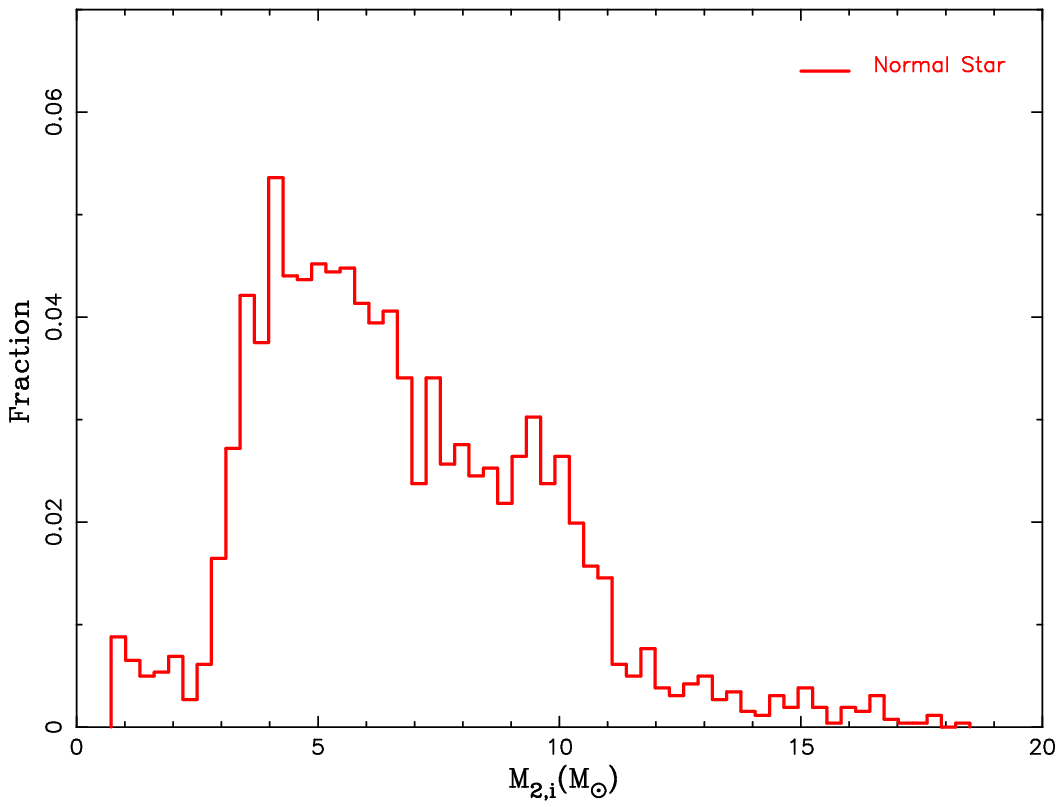}
\includegraphics[totalheight=2.85in,width=3.8in]{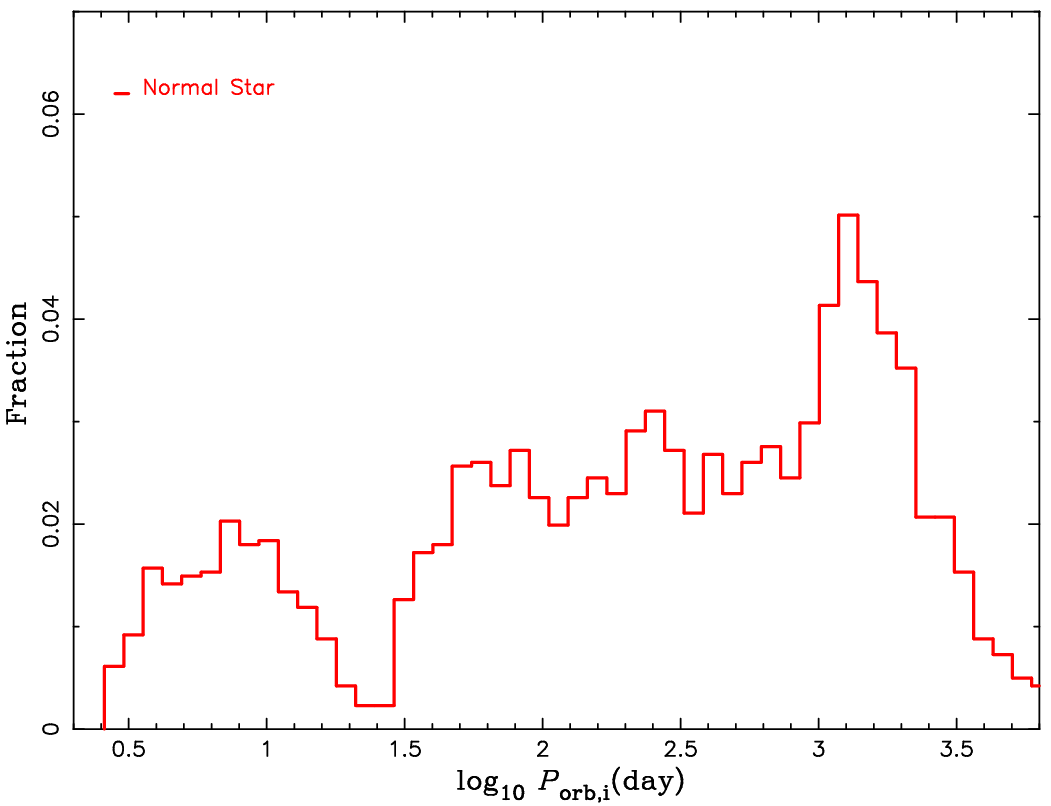}
\caption{The distributions of progenitors of ULX pulsar binaries at the beginning.
The initial masses of primary \& secondary stars, and initial orbital periods are
shown in upper, middle and lower panels, respectively.}
\end{figure}

\clearpage

\begin{figure}
\centering
\includegraphics[totalheight=2.9in,width=3.8in]{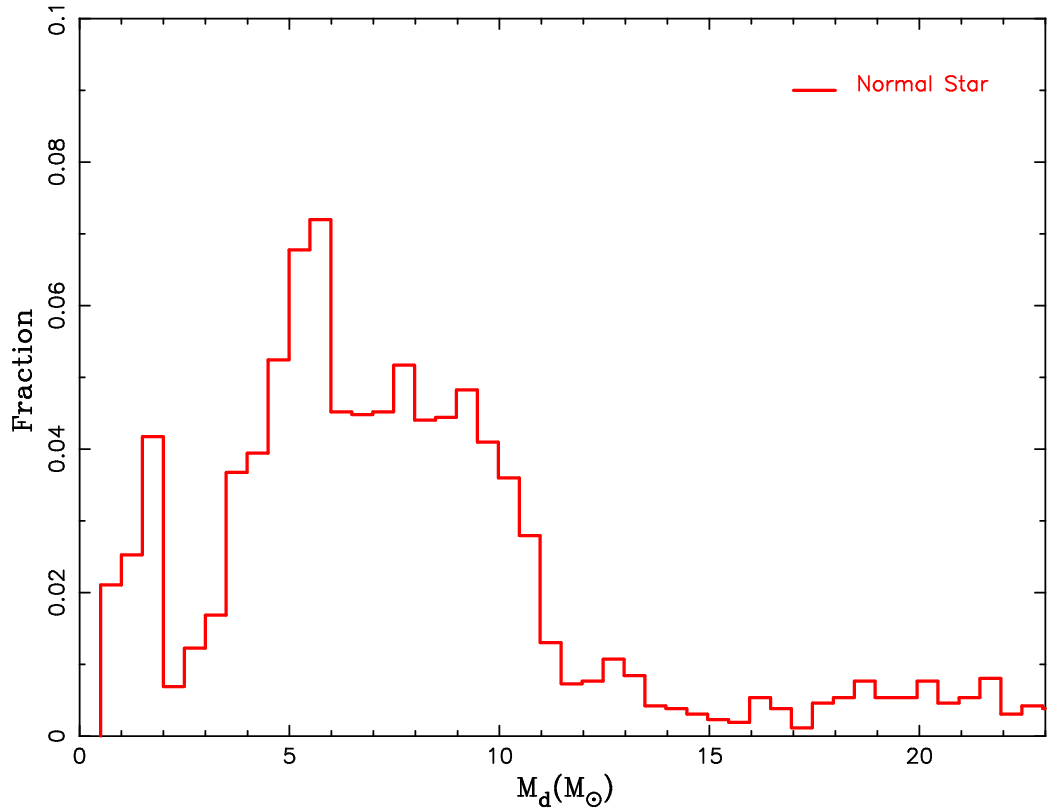}
\includegraphics[totalheight=2.9in,width=3.8in]{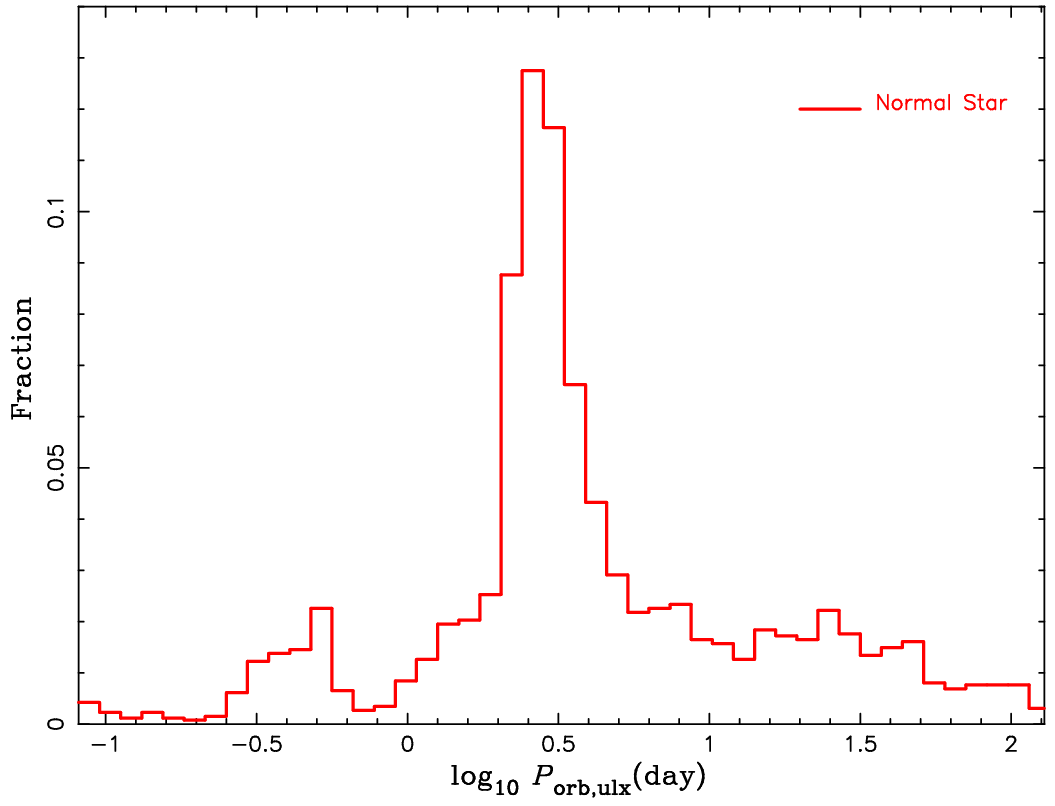}
\includegraphics[totalheight=2.9in,width=3.8in]{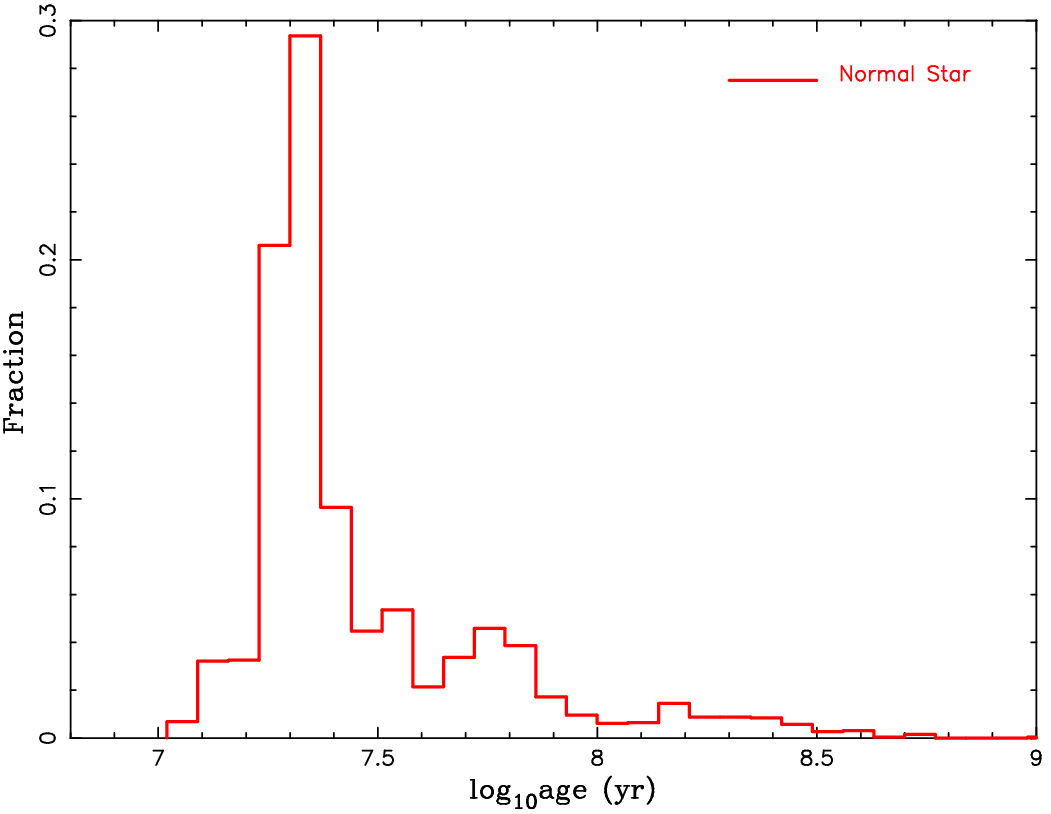}
\caption{The distributions of the donor mass, orbital periods and ages of ULX
pulsar binaries have given in upper, middle and lower panels, respectively.}
\end{figure}

\clearpage

\begin{figure}
\centering
\includegraphics[totalheight=4in,width=4.5in]{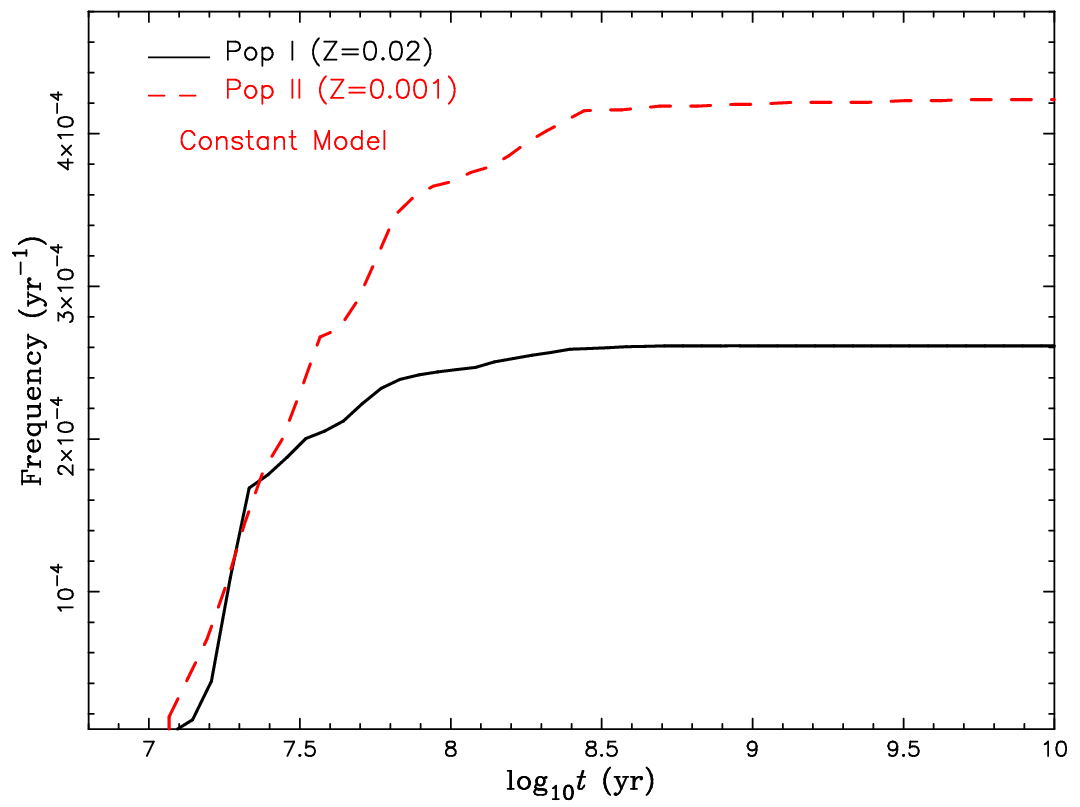}
\includegraphics[totalheight=4in,width=4.5in]{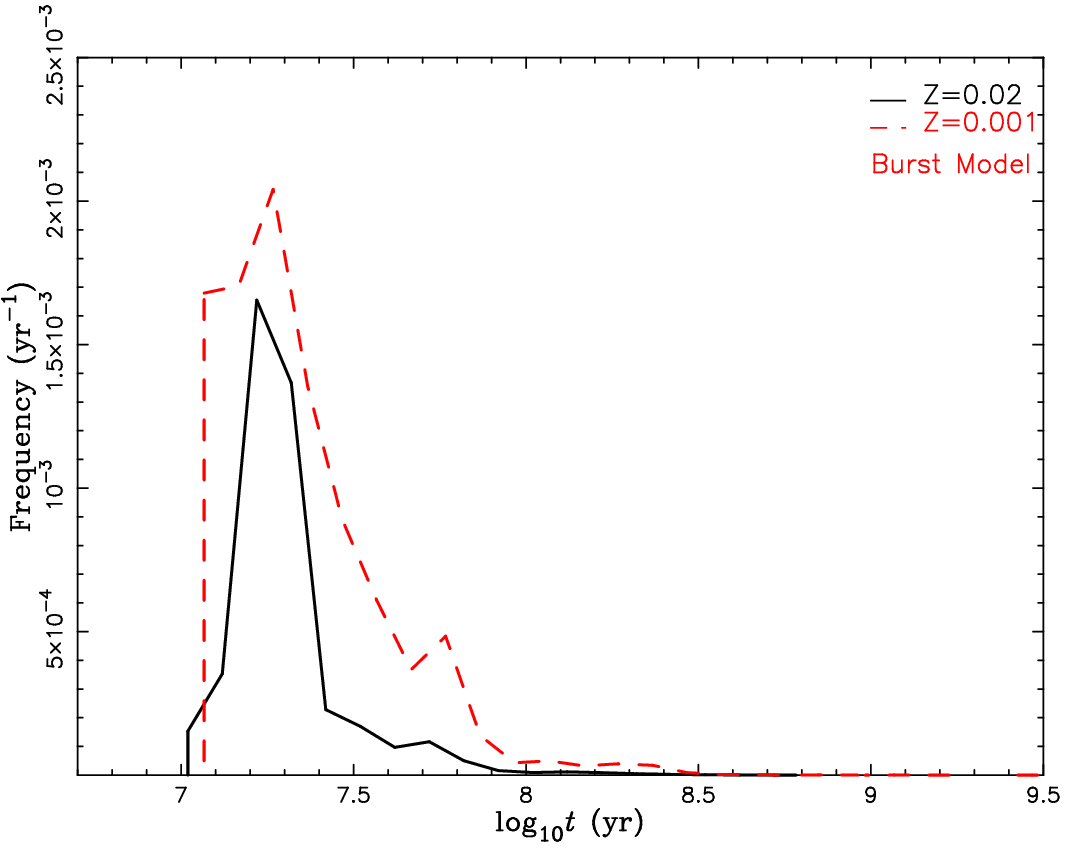}
\caption{The rate evolutions of ULX pulsar binaries under a constant star
formation model (upper) and single burst star formation model (lower).
}
\label{fig:1}
\end{figure}

\clearpage

\begin{figure}
\centering
\includegraphics[totalheight=3.3in,width=3.2in]{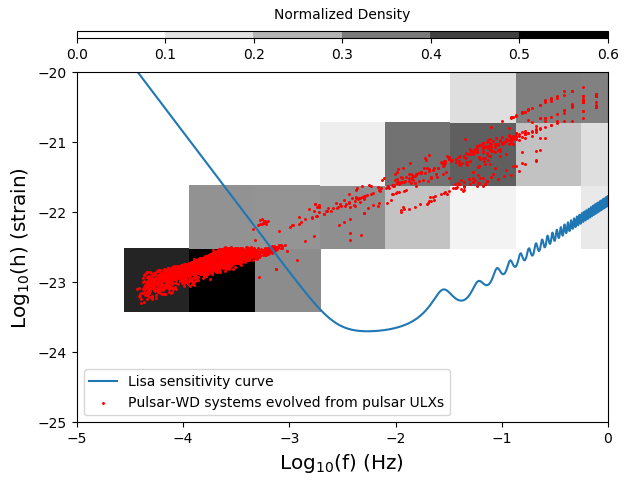}
\includegraphics[totalheight=3.3in,width=3.2in]{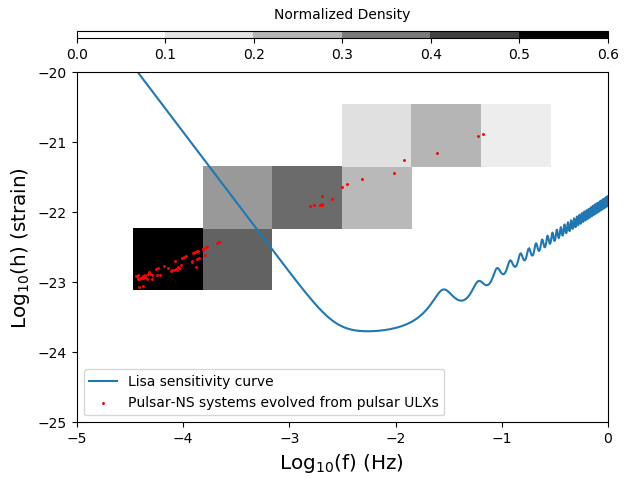}
\caption{The distribution of GW-sources from NS-WD (right) and NS-NS (left panel) mergers in the strain-frequency space. The sources are assumed to be at 10 kpc distances.
The sensitivity curve of LISA is shown with the solid blue line. Gray scale are all NS-WD and NS-NS sources in our simulation. Red dots are pulsar-WD and pulsar-NS sources
evolved from ULX pulsar binaries.
}
\label{fig:1}
\end{figure}

\clearpage

\end{document}